# Development of simple quantitative test for lack of field emission orthodoxy

RICHARD G. FORBES[*]

*Advanced Technology Institute, Faculty of Engineering and Physical Sciences,*

*University of Surrey, Guildford, Surrey GU2 7XH, UK*

This paper describes a simple quantitative test applicable to current-voltage data for cold field electron emission (CFE). It can decide whether individual reported field-enhancement-factor (FEF) values are spuriously large. The paper defines an "orthodox emission situation" by a set of ideal experimental, physical and mathematical conditions, and shows how (in these conditions) operating values of scaled barrier field ($f$) can be extracted from Fowler-Nordheim (FN) and Millikan-Lauritsen (ML) plots. By analyzing historical CFE experiments, which are expected to nearly satisfy the orthodoxy conditions, "apparently reasonable" and "clearly unreasonable" experimental ranges for $f$ are found. These provide a test for lack of orthodoxy. For illustration, this test is applied to 17 post-1975 CFE data sets, mainly for carbon and semiconductor nanostructures. Some extracted $f$-value ranges are apparently reasonable (including many carbon results), some are clearly unreasonable. It is shown that this test applies to any field-emission diode geometry and any form of FN or ML plot. It is proved mathematically that, if the extracted $f$-value range is "unreasonably high", then FEF-values extracted by the usual literature method are spuriously large. Probably, all new field-emitter materials should be tested in this way. Appropriate data-analysis theory needs developing for non-orthodox emitters.

[*]Author for correspondence (r.forbes@trinity.cantab.net)

## 1. Introduction

### (a) Motivation

Particular stimuli for this work have been the widespread interest in developing large area field emitters (LAFEs) for electronic device applications (e.g., Zhu 2001, Xu & Huq 2005, Saito 2010, Zhai et al. 2011), and the author's perception—shared informally with colleagues—that some reported values for the LAFE characterization parameter "field enhancement factor" are implausibly high. Thus, a specific aim is to improve and generalize recently reported tests (Forbes 2012a,b) that can decide whether the FEF value extracted from a given experimental data set is clearly unreliable.

The more general objective is a simple, well-specified quantitative test that can decide whether current-voltage characteristics measured in field electron emission experiments are incompatible with a set of experimental, physical and mathematical assumptions widely used in the period 1970 to 1990, and refined here to become the "orthodox emission hypothesis" described below. These are based around a particular form of Fowler-Nordheim-type equation. The test is intended for use on both modern and historical field emission data.

### (b) Background terminology

The following terms are used. *Fowler-Nordheim (FN) tunnelling* is electron tunnelling though an exact or approximately triangular barrier. *Deep tunnelling* is tunnelling well below the top of the barrier, in a regime where the Landau and Lifschitz (1956) expression for transmission probability is a valid approximation (see Forbes 2008c). *Cold field electron emission (CFE)* is the statistical emission regime in which the electrons in the emitting region are effectively in local thermodynamic equilibrium, and emission occurs mainly by deep FN tunnelling from states close in energy to the local emitter Fermi level.

Provided that any relevant emitting surface is "not too sharp" (radius of order 10 nm or greater), CFE from a metal conduction band is physically described by a family of approximate equations known as *Fowler-Nordheim-type (FN-type) equations* (see Forbes 2012a); different family members involve different levels of theoretical approximation and/or use different (but linked) dependent and/or independent variables. Some FN-type equations are also used as empirical fitting equations for other types of CFE, but the interpretation of extracted parameters may be problematic if the emission is not orthodox as described below.

*(c) Orthodox field electron emission*

At present, many details of CFE theory are incomplete, particularly for non-metals, and there are some general difficulties, particularly with the theory of current-voltage data interpretation. The definition of orthodox field electron emission provides statements of a "paradigm" body of basic CFE theory, and of the (ideal) conditions under which it can be validly applied to predict emitter behaviour and interpret measured current-voltage data. Emission from old-style metal single-tip emitters, of moderate to large tip radius, is expected to be orthodox, or nearly so. Real present-day emission situations may or may not correspond adequately to orthodox emission.

These ideal conditions concern (i) the behaviour of an operating field electron emitter in its whole environment, including the vacuum system and the electrical measurement circuit, and (ii) the physical assumptions made and mathematical models used when analyzing measured current-voltage data. A situation is deemed orthodox if all of the following apply:

(A) the voltage difference between the emitting regions and a surrounding counter-electrode (all parts of which are at the same voltage) can be treated as uniform across the emitting surface and equal to the measured voltage $V_m$;

(B) for a given voltage $V_m$, the measured current $i_m$ can be treated as controlled solely by CFE at the emitter/vacuum interface, and is not significantly influenced by any other feature of the measuring circuit;

(C) emission can treated as if it involves deep tunnelling through a Schottky-Nordheim (SN) barrier (Schottky 1914, Nordheim 1928, Forbes & Deane 2007);

(D) the measured current $i_m$ is adequately described by a SN-barrier-related FN-type equation in which the only quantities that depend (directly or indirectly) on the measured voltage $V_m$ are the barrier-form correction factor $v_F$ discussed below and any field-like or voltage-like variable in the equation; and

(E) the emitter local work-function is constant (and constant across the emitting surface), and has a value close to that expected.

This definition of orthodox emission allows slight departures from the strict physical assumptions made in mathematical modelling, but excludes many complications that can occur in real, present-day emission situations. Excluded complications include significant effects due to: Fermi-level drop in the measuring circuit (usually called "voltage drop"), or other forms of "saturation"; patch fields; leakage currents; field emitted vacuum space charge; current-induced changes in temperature; field penetration and band-bending; small-emitter effects related to field fall-off and/or quantum confinement (e.g., Qin et al. 2011); and field-related changes in emitter geometry or emission area or local work-function.

The specific term "orthodox emission" was introduced in Forbes 2012a. The underlying thinking is much older. What is given above is a more careful statement of the assumptions involved.

*(d) The reason to test for orthodoxy*

In the period 1970 to 1990, nearly all literature analyses of measured CFE current-voltage data in effect assumed emission to be orthodox, and used related data-analysis theory. However, if emission is not orthodox, then orthodox data-analysis methods may not work properly, and may generate spurious values for extracted physical parameters. Thus, a test for identifying non-orthodox experimental behaviour would be useful.

In the last 20 years, especially with work on LAFEs, mathematically simpler forms of data analysis have been used, based on the elementary FN-type equation (which is a simplified form of the equation originally given by Fowler & Nordheim in 1928). As discussed in Forbes 2012a, this approach can cause difficulties. In fact, it would be better to carry out analyses on the basis of the orthodox emission hypothesis—provided that one can first test for lack of orthodoxy. If a given FN plot fails an orthodoxy test, then in nearly all cases the simpler analysis based on the elementary FN-type equation would also be invalid.

A recent stimulus for developing an orthodoxy test has been a research tender document issued by an European Agency. The required research involved (amongst other things) a literature search for highly-efficient field electron emitters. It would be natural to take high reported field enhancement factor (FEF) as a measure of good emitter quality. However, non-orthodox experimental behaviour, when analyzed by orthodox theory, may generate a spuriously high FEF-value (see § 6). Thus, a risk exists that a literature survey of FEF-values would selectively choose spuriously high results, and hence return spurious research conclusions.

*(d) Previous test development*

Sophisticated tests of the applicability of FN-type equations to CFE from metal emitters have been made (Dyke & Trolan 1953, Dyke & Dolan 1956). But, historically, only one simple test has existed for judging compatibility between experimental current-voltage data and FN-type physical assumptions. This has been to make a FN plot (Stern et al. 1929) of the data, in the form $[\ln\{i_\mathrm{m}/V_\mathrm{m}^2\}$ vs $1/V_\mathrm{m}]$ or equivalent, or to make a related semi-logarithmic plot, such as a Millikan-Lauritsen (ML) plot (Millikan & Lauritsen 1928, Forbes 2009). Linearity or near-linearity of such plots has often been taken to imply that $i_\mathrm{m}$ could be *physically predicted* by (as opposed to mathematically fitted by) a FN-type equation applying at the emitter-vacuum interface.

In reality, this qualitative, "linear or nearly-linear" test is necessary but not sufficient, because other physical situations could create FN-plot linearity. (For example, linearity is expected when $i_\mathrm{m}$ is controlled by a FN-tunnelling-based conduction process in a substrate onto which emitters have been deposited.) The tests discussed below involve the slope of a FN or ML plot, and are superior because they are quantitative.

A "mid-range" test was proposed in Forbes 2012a and a "range-ends" test in Forbes 2012b. These tests were created for LAFEs, and were described in a restricted form related to FN plots of type $\ln\{J_M/F_M^2\}$ vs $1/F_M$], where $F_M$ is macroscopic field and $J_M$ is the macroscopic (i.e., LAFE-average) emission current density. In fact, similar tests can be applied to any type of FN plot, and hence to traditional single-tip field emitters (STFEs) and to cylindrical-wire emitters, as well as to LAFEs.

*(e) Objectives and structure*

This paper has three main objectives: (1) to fill out background theory (including the use of scaled equations), and formally prove that the new tests are applicable to any type of FN plot; (2) to improve the range-ends test by applying it to selected older experiments expected to be (nearly) orthodox; and (3) to illustrate use of the improved test by applying it to selected CFE experiments (including many on LAFEs) where orthodoxy cannot necessarily be expected.

The structure of this paper is as follows. Section 2 provides background theory, some of which is being presented in a definitive form for the first time. Section 3 discusses the principles behind tests for lack of orthodoxy. Section 4 applies the existing range-ends test (Forbes 2012b) to historical data, and uses the outcome to refine this test. Section 5 applies the improved test to published FN plots associated with various materials. Section 6 considers the link between failure of an orthodoxy test and the generation of spurious values for field enhancement factor. Section 7 provides a summary and discussion.

In accordance with the mainstream field emission convention, this paper treats fields, current densities and related quantities as positive, even though they would be negative in classical electromagnetism, and uses $F$ as the symbol for the negative or magnitude of electrostatic field as classically defined.

## 2. Background theory

*(a) CFE equations for emission current densities*

This Section presents a coherent set of general FN-type equations. A formal general expression $J_L^{GB}$ for the *local emission current density (ECD)*, as a function of local work-function $\phi_L$ and local barrier field $F_L$, is (Forbes 2012a):

$$J_L^{GB} \equiv \lambda_L^{GB} a \phi_L^{-1} F_L^2 \exp[-v_F^{GB} b \phi_L^{3/2}/F_L], \tag{1}$$

where $a$ and $b$ are the First and Second FN Constants as given in Table 1, $v_F^{GB}$ (" $\text{nu}_F^{GB}$ ") is a particular value (for a barrier of zero-field height $\phi_L$) of the *barrier-form correction factor* $v^{GB}$ for a defined general mathematical barrier "GB", and $\lambda_L^{GB}$ is the corresponding *local pre-exponential correction factor*. $\lambda_L^{GB}$ is a composite correction factor that can in principle be decomposed into a product of correction factors related to specific physical effects, as in Forbes 2008b; details are not needed here.

In equation (1), the superscript "GB" appears on all relevant parameters, to show that their values depend on the mathematical form chosen to model the tunnelling barrier. Predictions of $\lambda_L^{GB}$ and $J_L^{GB}$ also depend on other assumptions made, but these are not shown explicitly.

In practice, both with old-style single-tip field emitters and with multi-emission-site large-area field emitters, theoretical interest is in the characteristic local current density $J_C$ at some characteristic location "C" on the emitter surface. It is helpful to think of "C" as the location at which $J_L^{GB}$ has its maximum value for a given applied voltage, but other choices are possible. Quantities applicable to location "C" are subscripted "C".

For discussing data analysis, it is useful to define a characteristic *kernel current density* $J_\text{C}^\text{kGB}$ for a general barrier, and re-write expression (1) (but with "L" replaced by "C") as the linked equations

$$J_\text{C}^\text{kGB} \equiv a\phi_\text{C}^{-1} F_\text{C}^2 \exp[-v_\text{F}^\text{GB} b\phi_\text{C}^{3/2}/F_\text{C}], \tag{2a}$$

$$J_\text{C}^\text{GB} = \lambda_\text{C}^\text{GB} J_\text{C}^\text{kGB}. \tag{2b}$$

This approach has two advantages. First, for given choices of $\phi_\text{C}$, $F_\text{C}$ and barrier form, values for the mathematical expression $J_\text{C}^\text{kGB}$ can be calculated exactly, because all the uncertainties in the pre-exponential (see Forbes 2008b) have been accumulated into the single parameter $\lambda_\text{C}^\text{GB}$. Second, this approach provides formal similarity between expression (2b) above, for the characteristic *local* ECD, and expression (3) below, for the *macroscopic (or "LAFE-average") ECD* $J_\text{M}$:

$$J_\text{M} = \lambda_\text{M}^\text{GB} J_\text{C}^\text{kGB}. \tag{3}$$

The *macroscopic pre-exponential correction factor (for the general barrier)*, denoted here by $\lambda_\text{M}^\text{GB}$, is the parameter introduced in Forbes 2012a and denoted by $\lambda_\text{M}$ in equation (11) there. The formal similarity between equations (2b) and (3) ensures that discussion below is applicable to both STFEs and LAFEs.

No superscript is used on $J_\text{M}$ because it is not model dependent but is defined operationally, using the total emission current $i$ and the LAFE substrate (or "macroscopic") area $A_\text{M}$, by

$$J_\text{M} = i/A_\text{M}. \tag{4}$$

Obviously, in any given experiment, both $A_\text{M}$ and $i$ can be measured. In LAFE theory based on a general barrier, the *area efficiency of emission* $\alpha_\text{M}^\text{GB}$ is given by

$$\alpha_\mathrm{M}^\mathrm{GB} \equiv J_\mathrm{M}/J_\mathrm{C}^\mathrm{GB} = \lambda_\mathrm{M}^\mathrm{GB}/\lambda_\mathrm{C}^\mathrm{GB} . \tag{5}$$

The above equations relate to current densities. For all emitters, equations relating to total emission current can be written in the form:

$$i \equiv A_\mathrm{n}^\mathrm{GB} J_\mathrm{C}^\mathrm{GB} , \tag{6}$$

where $A_\mathrm{n}^\mathrm{GB}$ is defined by equation (6) and is a *notional emission area (for the general barrier)* for the emitter in question. For LAFEs, the related form $i=A_\mathrm{M}J_\mathrm{M}$ can also be used.

*(b) Classification of Fowler-Nordheim–type equations by sub-family*

The kernel current density $J_\mathrm{C}^\mathrm{kGB}$ has a central role in the above equations. This suggests splitting the family of FN-type equations into sub-families based on the tunnelling barrier model used. Each sub-family has a defined expression for the kernel current density; the different equations in the sub-family then use different approximate expressions for the pre-exponential correction factor (or use different sets of dependent and independent variables).

At present, most CFE data analyses are based on FN-type equations drawn either from the sub-family based on the exact triangular (ET) barrier (for which $v^\mathrm{GB} = 1$), or from the sub-family based on the SN barrier (for which $v^\mathrm{GB}$ is given by a particular value of the principal SN-barrier function $v(l')$, where $l'$ is a mathematical variable (Deane & Forbes 2008)). The ET barrier leads to slightly simpler mathematics, but the SN barrier is physically and quantitatively more realistic. The elementary FN-type equation used in many analyses of LAFE behaviour is the simplest member of the sub-family based on the ET barrier. But, in the context of LAFEs, the elementary equation is seriously defective (see Forbes 2012a), in that it omits any macroscopic pre-exponential correction factor, and hence greatly over-predicts $J_M$ (by a factor that could in some cases be as much as $10^9$ or more).

For data analysis, the SN-barrier-based sub-family is much the best choice at this stage of theoretical development. The "full" SN-barrier-based equation for a LAFE, namely

$$J_M^{SN} = \lambda_M^{SN} J_C^{kSN}, \quad \text{with} \quad J_C^{kSN} \equiv a\phi_C^{-1} F_C^2 \exp[-v_F b\phi_C^{3/2}/F_C], \tag{7}$$

is central to the concept of emission orthodoxy.

In due course, if it emerges that many LAFEs do not exhibit orthodox emission, it may become useful to develop data-analysis theory based on other barrier models. At present, it seems best to concentrate on consolidating data-analysis theory for orthodox emission, since gaps in this still exist.

*(c) Scaled form for the Schottky-Nordheim kernel current density*

This Section records how expression (7) can be converted to "scaled" form. Scaled forms allow arguments to be made wider in their applicability, and SN-barrier mathematics is particularly amenable to scaling. The discussion here improves on that given in the appendix in Forbes 2012a.

For a SN barrier of zero-field height $\phi$, the *reference field* $F_R^{SN}$ that reduces the barrier height to zero is:

$$F_R^{SN} = c^{-2}\phi^2, \tag{8}$$

where $c$ is the Schottky constant, as given in Table 1. For this barrier, the *scaled barrier field* $f^{SN}$ corresponding to barrier field $F$ is

$$f^{SN} \equiv F / F_R^{SN}. \tag{9}$$

$f^{SN}$ is a dimensionless physical quantity used in modelling the SN barrier. The process of deriving equation (7) shows that $v_F$ can be identified with the particular value $v(f_C^{SN})$ obtained by putting $l' = f_C^{SN}$ in the mathematical function $v(l')$.

On defining work-function-dependent parameters $\eta^{SN}(\phi)$ and $\theta^{SN}(\phi)$ by

$$\eta^{SN}(\phi) = b\phi^{3/2}/F_R^{SN} = bc^2\phi^{-1/2}, \tag{10}$$

$$\theta^{SN}(\phi) = a\phi^{-1}(F_R^{SN})^2 = ac^{-4}\phi^3, \tag{11}$$

where $bc^2$ and $ac^{-4}$ are universal constants given in Table 1, the kernel current density $J_C^{kSN}$ can be written in scaled form as

$$J_C^{kSN} = \theta_C^{SN} \cdot (f_C^{SN})^2 \cdot \exp[-v(f_C^{SN}) \cdot \eta_C^{SN}/f_C^{SN}]. \tag{12}$$

Values of $\eta^{SN}(\phi)$ and $\theta^{SN}(\phi)$ are shown in Table 2, for a range of work-function values.

To make equations physically explicit, the subscript "C" has been used above to label characteristic values, and superscripts have been used to label different barrier models. For notational simplicity, we now drop the tag "SN" entirely, and drop the subscript "C" from the parameters $\phi_C$, $F_C$, $f_C^{SN}$, $\eta_C^{SN}$, $\theta_C^{SN}$ and $J_C^{kSN}$, leaving it to be understood that from this point—unless indicated otherwise—relevant parameters relate to SN-barrier-based equations for location "C".

The scaled equation (12) contains only one field-like variable (*f*), and a good simple approximation exists for $v(f)$ (Forbes & Deane 2007, 2010). This allows the SN-barrier kernel current density $J^k$ to be well approximated by

$$J^k \approx \theta f^2 \exp[-\eta \cdot (1/f - 1 + \tfrac{1}{6}\ln f)]. \tag{13}$$

For example, for $\phi$=4.50 eV and $f$=0.30, the exponent in equation (13) evaluates to –9.889, as compared with an unapproximated value of –9.856, and $J^k$ evaluates to 3.20×10$^8$ A/m$^2$, as compared with an unapproximated value of 3.09×10$^8$ A/m$^2$.

*(d) Applicability of tests to all forms of Fowler-Nordheim plot*

This Section shows that the new tests apply to all forms of FN plot. It is helpful to introduce a formal mathematical differential operator $\mathcal{D}_X\{Y(X,Z)\}$. This operates on any well-behaved function $Y(X,Z)$ that is a function of one variable ($X$) of particular interest, and possibly of a set $Z$ of other variables that are independent of $X$ and not of immediate interest, and is defined by

$$\mathcal{D}_X\{Y(X,Z)\} \equiv [\partial \ln\{Y\}/\partial(1/X)]_Z . \tag{14}$$

Clearly, for two functions $Y_1(X,Z)$ and $Y_2(X,Z)$:

$$\mathcal{D}_X\{Y_1 \cdot Y_2\} = \mathcal{D}_X\{Y_1\} + \mathcal{D}_X\{Y_2\} . \tag{15}$$

For any parameter $K$ that is independent of $X$:

$$\mathcal{D}_X\{K\} = 0 . \tag{16}$$

This operator is first employed to investigate the use of different *dependent* variables in FN plots. A plot of type [ln{$Y/X^2$} vs 1/$X$] is said to be "plotted using FN coordinates for $Y(X)$" and is called a "$Y$-$X$" FN plot or $Y(X)$ FN plot. When a function $Y(X)$ is plotted as a FN plot, the plot slope is denoted by $S_{YX}(X)$. For FN-type equations this slope is negative, and varies slightly with $X$ (except for the ET-barrier sub-family).

In orthodox emission, it is assumed that $A_M$, $A_n$, $\lambda_M$, $\alpha_M$, $\lambda_C$, $\phi$, $\theta$ and $\eta$ can all be treated as independent of fields and voltages. Thus, from equation (12), the predicted slope $S_{kf}$ of a $J^k$-$f$ FN-plot would be

$$S_{kf} = \mathcal{D}_f\{J^k/f^2\} = \mathcal{D}_f\{\theta\} - \eta \cdot d\{v/f\}/d(1/f) = -\eta \cdot s(f), \qquad (17)$$

where $s(f)$ is the slope correction function for the SN barrier, expressed in terms of $f$. The mathematical function $s(f)$ is defined and well-approximated by (Forbes & Deane 2007)

$$s(f) \equiv d\{v/f\}/d(1/f) = v - f dv/df \approx 1 - f/6. \qquad (18)$$

For the SN-barrier sub-family, using the simplified notation just defined, equation (2b) takes the form $J_C = \lambda_C J^k$. In orthodox emission $\mathcal{D}_f\{\lambda_C\}=0$. Thus, the predicted slope $S_{Cf}$ of a $J_C$-$f$ FN-plot is

$$S_{Cf} \equiv \mathcal{D}_f\{J_C/f^2\} = \mathcal{D}_f\{\lambda_C\} + \mathcal{D}_f\{J^k/f^2\} = -\eta \cdot s(f). \qquad (19)$$

Obviously, this result is the same as in equation (15). Analogous arguments, starting from the relation $J_M = \lambda_M J^k$, prove that the slope $S_{Mf}$ of a theoretical $J_M$-$f$ FN-plot would also be given by $-\eta \cdot s(f)$.

Equation (6) records that, for all emitters, the total emission current $i$ can be written as $i=A_n J_C$, where $A_n$ is the relevant notional emission area. Orthodox emission treats $A_n$ as constant; hence $\mathcal{D}_f\{A_n\}=0$, and the predicted slope $S_{if}$ of an $i$-$f$ FN plot is again $-\eta \cdot s(f)$.

In what follows, the symbol $S_{Yf}$ is used to represent any one of $S_{kf}$, $S_{Cf}$, $S_{Mf}$ or $S_{if}$. Clearly, in all cases $S_{Yf} = -\eta \cdot s(f)$.

The $\mathcal{D}$-operator can also be used to investigate what happens when different *independent* variables are used in FN plots. The case of "voltage" can be taken as an example. In diode-like practical situations, where the emitter faces surrounding surfaces all at the same voltage, the measured independent variable is usually the voltage-difference $V_m$ supplied by the high-voltage generator. In orthodox emission, all emitting surfaces are assumed to be at the same voltage, and the gap voltage-difference $V_G$ (between the emitter and its surroundings) is treated as equal to $V_m$. The symbol $V$ is used here to denote these equal voltage-differences, and $V_R$ to denote the (constant) reference value of $V$ at which (for the given emitter, with a defined atomic-level shape) a SN barrier of zero-field height $\phi$, at location "C", is reduced to zero. For this emitter shape, orthodox emission assumes that corresponding values of $V$ and $f$ are related exactly by

$$V = f \cdot V_R, \tag{20}$$

$$d(1/V)/d(1/f) = 1/V_R; \tag{21}$$

and, from definition (14), it follows (for any function $Y$) that

$$\mathcal{D}_V\{Y\} = \mathcal{D}_f\{Y\} \cdot \frac{d(1/f)}{d(1/V)} = \mathcal{D}_f\{Y\} \cdot V_R. \tag{22}$$

Consider a set of theoretically prepared CFE data (say, a set of pairs of values of $f$ and $J^k$) that is plotted both as a $J^k$-$f$ FN plot, and (after transformation using some specific $V_R$-value) as a $J^k$-$V$ FN plot. The predicted slopes $S_{kV}$ and $S_{kf}$ of these plots will—for corresponding values of $V$ and $f$—be mathematically related. This can be formally proved as follows.

$$S_{kV} = \mathcal{D}_V\{J^k/V^2\} = \mathcal{D}_V\{J^k/(f^2 V_R^2)\}$$

$$= \mathcal{D}_V\{J^k/f^2\} + \mathcal{D}_V\{1/V_R^2\} = \mathcal{D}_V\{J^k/f^2\}$$

$$= \mathcal{D}_f\{J^k/f^2\} \times \mathrm{d}(1/V)/\mathrm{d}(1/f) = S_{kf}/V_R = (S_{kf}/f)\cdot V, \quad (23)$$

$$S_{kV}/V = S_{kf}/f.$$

A similar proof could be given for each of the other dependent quantities ($J_C$, $J_M$ and $i$) considered above, and the set of results can be summarised in the statement: $S_{YV}/V = S_{Yf}/f = -\eta \cdot s(f)/f$.

Since, in orthodox emission conditions, the relationships between $f$ and barrier field $F$, and between $f$ and macroscopic field $F_M$, also involve strict proportionality, similar conclusions apply to theoretical FN plots using $F$ and $F_M$ as the independent variables. Thus, if $S_{YF}$ and $S_{YM}$ represent the corresponding sets of FN-plot slopes, we would find that $S_{YF}/F$ and $S_{YM}/F_M$ again equal $-\eta \cdot s(f)/f$.

What this Section has shown is the following. If $X$ is any one of the independent variables normally used in CFE theory, and $Y$ is any one of the dependent variables, then under conditions of emission orthodoxy the following is true: that, at any particular value of $X$, the ratio $S_{YX}/X$ is given by

$$S_{YX}/X = -\eta \cdot s(f)/f, \quad (24)$$

where $f$ is the value of scaled barrier field corresponding to the chosen value of $X$.

This result makes it possible to apply quantitative orthodoxy tests (which use the variable $f$) to any type of FN plot. The discussion also brings out that, when emission is orthodox, $f$ can be used as a scaled variable for any of the independent variables $F$, $V$ or $F_M$, by choosing a reference value at which the barrier height goes to zero. (This is not necessarily true for non-orthodox situations, because lack of orthodoxy may involve breakdown in strict proportionality between $F$ and other independent variables.)

An alternative (less formal) derivation of result (24) is given in Forbes et al., 2012.

### 3. Quantitative tests for lack of emission orthodoxy

*(a) Extraction of f-values*

With FN-type equations based on the SN barrier, all experimental FN plots (of whatever type) are expected to be slightly curved (e.g., Forbes & Deane 2010). A straight line of slope $S^{\mathrm{fit}}$, fitted to data points by linear regression or other means, is a chord to the theoretical FN plot generated by the FN-type equation being used to model the data. This fitted line is parallel to the tangent to the model FN plot at some value $X_t$ of the independent variable $X$. Hence, if $f_t$ corresponds to $X_t$ (i.e., $f_t = X_t/X_R$, where $X_R$ is the reference value for variable $X$), then

$$S^{\mathrm{fit}}/X_t = -\eta \cdot s(f_t)/f_t , \tag{25}$$

$$S^{\mathrm{fit}}/X = -\eta \cdot s(f_t) \cdot X_t/f_t X = -\eta \cdot s(f_t)/f , \tag{26}$$

$$f^{\mathrm{extr}} = \{\eta \cdot s(f_t)\} / \{-S^{\mathrm{fit}} \cdot (1/X)^{\mathrm{expt}}\} . \tag{27}$$

The last step in equation (26) follows because (for orthodox emission) $X_t/f_t = X_R$ and $X_R/X = 1/f$. Equation (27), obtained by inverting equation (26), shows how an extracted value $f^{\mathrm{extr}}$ is obtained from the experimental value $(1/X)^{\mathrm{expt}}$ read from the horizontal $(1/X)$ axis of a FN plot.

A value for $s(f_t)$ is now needed. A good simple approximation exists for $s(f)$, namely $s(f) \approx 1 - f/6$. From re-analysis (Forbes 2008a) of the experimental results of Dyke & Trolan (1953) for their emitter X89, it was found that the $f$-value range corresponding to their results is $0.20 < f < 0.34$. The corresponding range for $s(f)$ is $0.967 > s > 0.943$. Hence, to an adequate approximation, since $f_t$ is near the middle of the stated range, one may take $s(f_t) \approx 0.95$. This was the approach adopted in the range-ends test of Forbes 2012b.

The Dyke & Trolan results for emitter X89 were used because this was the main set of results used in the 1950s tests of the applicability of FN theory to CFE data. Emitter X89 is a very carefully investigated tungsten single-tip field emitter.

In the original mid-range test, a different method was used to deal with the $s(f_t)$ term. In effect, $s(f_t)$ was expanded as $(1–f_t/6)$, a working value $X_w$ was chosen near the midpoint of the range of $1/X$ used in the relevant experiments (or simulations), and it was assumed that $f_t$ could be approximated as equal to the scaled barrier field $f_w$ corresponding to $X_w$. This approach attempts to make $s(f_t)$ consistent with the experimental data, but can yield unphysical values for $s(f_t)$ when emission is not orthodox. The approach of approximating $s(f_t)$ as 0.95 makes $s(f_t)$ consistent with tunnelling theory, and is considered simpler and better, for both the mid-range and range-ends tests.

*(b) Formulation of test criteria*

To create a test, one needs to know what range of *f*-values is physically reasonable (usually this means reasonable for steady emission). An *f*-value extracted from a linear (or nearly linear) experimental FN plot is then considered inconsistent with orthodox emission if it lies significantly outside the physically reasonable range.

The original mid-range test took $0.22 < f_w < 0.32$, which was thought reasonable for the extracted value $f_w$ corresponding to a "working value" $X_w$ (of the independent variable $X$), chosen near the midpoint of the experimental range of $1/X$ values. The range-ends test developed later used the whole range of extracted *f*-values, and compared this with the assumed physically reasonable range $0.20 < f^{extr} < 0.34$. (A modified range is proposed below.)

When the whole extracted range is used, the test criterion is as follows. If the FN plot is linear or very nearly linear, then inconsistency with orthodox emission is demonstrated if *any value in* the extracted range is significantly outside the physically reasonable range. (What "significantly" should mean is considered below.)

## 4. Application to older CFE experiments

*(a) Results of analysis*

Old-style CFE experiments, both those involving a metal single-tip field emitter mounted on a metal support and older experiments involving a metal wire and a concentric metal cylinder, are expected to exhibit orthodox (or nearly orthodox) behaviour. For some of the best-known CFE experiments between 1926 and 1972, extracted *f*-value ranges are shown in Table 3. In all cases the work function is assumed to be 4.5 eV.

Data set 1 is that used earlier, and is also shown as Fig.13 in Gomer's (1961) well-known textbook. Data set 6 is also shown as Fig. 8 in Good & Müller's (1956) review article. Data set 10, taken in wire-and-concentric-cylinder geometry (see Eyring et al. 1928), is the current-voltage ($i_m$-$V_m$) data set with which Lauritsen first discovered linearity between $\log_{10}\{i_m\}$ and $1/V_m$ (see Millikan & Lauritsen 1928, Lauritsen 1929, Holbrow 2003). Results in Rother's 1926 review are excluded because, for the tip-anode spacing tested in his Tafel 5 (0.015 mm), a FN re-plot of his results does not yield a straight line. Further information about the experimental data sets used to produce Table 3, and related references, and details of the related spreadsheet calculations, are provided as electronic supplementary material.

In the 90 years of practical CFE since the first report in English (Lilienfeld 1922) of what was then called auto-electronic emission, Table 3 appears to be the first published attempt to make a systematic *quantitative* comparison of classical CFE experimental results. The overall picture is one is of general consistency.

*(b) Refinement of orthodoxy test*

Clearly, many experiments used slightly lower onset conditions than were used for emitter X89; hence a revised test lower limit will be set at $f^{extr}$=0.15. Data set 8 is the only one with a still lower onset (0.11). This could be because Abbott and Henderson (1939) used an electrometer for their low-current measurements, rather than the galvanometer used for their higher-current measurements.

For steady emission, data sets taken in the 1950s and later terminate (on the high-$f$ side) at 0.35 or less. However, the older data sets go up as far as $f$ = 0.59. This difference could be due to increased later awareness that emitting high steady current density can cause premature emitter destruction. Another factor may be that two data sets involving high $f$-values (Nos 7 and 11) were parts of experimental investigations of the effects of temperature on emission, where it would be natural to use relatively high field values in the room-temperature data sets. Given the historical spread, a revised test upper limit for modern results is set, somewhat arbitrarily, at $f^{extr}$=0.45.

Another table feature, where pulsed emission has been used to explore emission at higher $f$-values, is the variability in the $f$-value at which non-linearity is detected. This could partly be due to difficulties in quantifying the onset of non-linearity; alternatively, the most obvious physical explanation lies in geometrical differences in space-charge effects.

Because of the difficulties in setting precise test limits before gaining experience in its practical use, the range-ends test currently needs to be treated as an "engineering triage" test: it sorts experimental results into the three categories "apparently reasonable", "clearly unreasonable" and "more detailed investigation required". The limits proposed above are the limits of the central "apparently reasonable" range (for an emitter with $\phi$= 4.5 eV). As boundaries for the "clearly unreasonable" range, the author proposes $f^{extr}$<0.10 and $f^{extr}$>0.75, the former because it is slightly lower than any observed onset value, the latter because it is somewhat higher than any $f$-value used in the pulsed-emission experiments listed.

These proposed limits may need adjustment in the light of experience and/or for certain classes of emitter. For example, it is arguable that very sharp emitters—including liquid-metal field electron emitters—might be able to sustain current densities corresponding to $f$-values closer to unity.

*(c) Tests for other work-function values*

For "orthodoxly behaving" materials with work-function values different from the tungsten value, both the value of $\eta$ and the ranges of "apparently reasonable" ($f_{low.} \leq f^{extr} \leq f_{up}$) and "clearly unreasonable" ($f^{extr} < f_{lb}$, and $f^{extr} > f_{ub}$) $f$-values will be different. It seems probable that the effects determining the physically reasonable emission range relate strongly to the emitter's characteristic local emission current density $J_C$. This, in turn, relates to the kernel current density $J^k$. For $\phi$=4.50 eV, the range-boundary values ($J_{lb}$, $J_{low.}$, $J_{up.}$, $J_{ub}$) are, respectively 3.02×10$^{-6}$, 25.6, 8.79×10$^{10}$ and 1.02×10$^{13}$ A/m$^2$. For other work-function values, the boundary $f$-values can be found by trial-and-error methods based on equation (12). These values are shown in Table 2, and can be used to provide an orthodoxy test for a material with work function different from 4.50 eV.

## 5. Illustrative applications to post-1975 data sets

This modified test is now applied to selected examples of various emitting materials. Table 4 shows the material, the work-function assumed and extracted $f$-values. Values outside the "apparently reasonable" range are marked with a single asterisk, "clearly unreasonable" values with a double asterisk. Where given by the original author, the extracted "field enhancement factor" (parameter $\gamma^{*1}$ in equation (28) below) is also shown. Details of table preparation are provided as electronic supplementary material. Brief notes on individual tests follow.

Test 15 relates to a Spindt array (Spindt et al., 1976). The upper end is outside the apparently reasonable range. This could be because the emitters were being operated at particularly high current density, or because the individual emitters have small tip radius, or a mixture of both.

Test 16 relates to a hybrid nanostructure that uses relatively closely spaced gold nanowires grown on a graphene film. The whole extracted range is "clearly unreasonable". This result has already been reported (Forbes 2012b, Lee et al. 2012), and is not fully understood. Possible explanations include field-dependent changes in emitter geometry and/or collective screening effects.

Tests 17 and 18 relates to what the authors (Li et al., 2011) call a "flexible $SnO_2$ nanoshuttle". The related FN plot has two quasi-linear sections. For both sections, the whole extracted range is "clearly unreasonable"; this could be an effect related to field-dependent geometry.

Tests 19 to 23 relate to various forms of carbon field emitter. Apart from two end-values, the whole extracted range is "apparently reasonable" in every case. Carbon field emission is not necessarily well described by FN-type equations, so this result is slightly unexpected. It may indicate that these carbon emitters (and their supporting structures) have adequate conductivity, and that carbon barrier behaviour is sufficiently close to that of a SN barrier.

Test 21 is particularly interesting, since it relates to a large-scale electrical engineering device, namely a gas-discharge tube surge protector (used, for example, to protect telecommunication circuits from lightning strikes). The authors (Zumer et al., 2012) consider that gas breakdown is initiated/stimulated by CFE from the edge of graphite platelets used in the device: obviously, the test result here supports this proposition.

Tests 24 and 25 relate to a CNT mat on a silicon substrate. The FN plot has two sections, with the high-field one obviously a "saturated" regime. The two tests between them clearly pick out the high-$f$ regime as non-orthodox.

Tests 26 and 27 relate to two different CNT-in-matrix composites. The first is "apparently reasonable", the other not. A point of interest is that the FN plot used for test 27 has only a single quasi-linear section: there is no "unsaturated" section.

Tests 28 to 31 relate to four different kinds of semiconductor nanostructure. For tests 29 to 31, the FN plots are chosen because they have the three highest "field enhancement factors" listed in Table 1 in the review by Zhai et al. (2011). As can be seen, one of these is "apparently reasonable", but the other three are non-orthodox. This means that one of the four high reported values of "field enhancement factors" is apparently valid, but the other three are apparently spurious (see § 6).

The comments provided here are not intended as definitive scientific conclusions. Rather, the tests and comments are intended as "proof of concept" and as a "demonstration of usefulness". They illustrate the kinds of issues that applying the test might raise or illuminate.

## 6. Saturation and spuriously large values for field enhancement factors

A direct mathematical link exists between (a) non-orthodoxy associated with current saturation, and (b) spuriously large values of field enhancement factor (FEF). A formula written here as

$$\gamma^{*1} = -b\phi^{3/2} / S^{\text{fit}} \tag{28}$$

is widely used to derive an experimental value for a *slope characterization parameter* denoted here by $\gamma^{*1}$ (but usually by $\beta$ in LAFE literature). This parameter $\gamma^{*1}$ is often interpreted as a FEF. However, if, due to saturation, the magnitude $|S^{\text{fit}}|$ of the FN plot slope is significantly lower than the value predicted for orthodox emission, then this anomalously low value will generate both a spuriously large FEF-value via equation (28) and anomalously high extracted *f*-values in the test based on equation (27).

For safety of interpretation, all experiments for which equation (28) generates high FEF-values need to be checked for lack of orthodoxy. As discussed in §5, Table 4 shows that some reported high FEF-values are spurious.

A formal improvement denotes the classical electrostatic (zero-current) field enhancement factor by the separate symbol $\gamma^{*\sigma}$. The underlying problem is that, if equation (28) is treated as an empirical extraction formula for this FEF, then the extraction formula is defective. A formal improvement writes it as

$$\gamma^{*\sigma} = \sigma_t \gamma^{*1} = -\sigma_t b \phi^{3/2} / S^{\text{fit}}, \tag{29}$$

where $\sigma_t$ is a slope correction factor. For orthodox emission $\sigma_t = s(f_t) \approx 0.95$, as in equation (27), so approximating $\sigma_t \approx 1$ causes only about 5% error. For non-orthodox emission, $\sigma_t$ may be very different from unity, and needs to be estimated theoretically. This is not straightforward, because each of the many possible causes of non-orthodoxy may need to be considered separately.

A further difficulty also exists. Even if the orthodoxy test indicates that a large extracted FEF-value is "apparently reasonable", this does not necessarily mean that the emitter nanostructure in question is technologically a "good field enhancer". The FEF-value is a reflection of the electrostatics of the whole system. Reports of continuous increase in FEF with cathode-anode separation (e.g., Lin et al. 2010) appear to indicate that, in some cases, effects due to the finite (and, maybe, unequal) sizes of electrodes may be influencing reported FEF-values. (This is an effect different from the better-known effect discussed by H. C. Miller (1967), which occurs at small tip-anode separations, for electrodes of effectively infinite extent.)

Detailed discussion of these issues is beyond the scope of the present paper. However, I urge again that, if FN plots are made, then these should involve the raw measured current-voltage (*i-V*) data, rather than data pre-converted (often, in past literature, by methods that are not fully described) to be in terms of $F_M$ and $J_M$. Issues relating to field enhancement factors (however these are defined) can then be discussed by multiplying the voltage-to-barrier-field conversion factors (extracted from the *i*(*V*) FN-plot slope) by appropriate well-defined distances. This approach of reporting (and working with) the raw experimental data seems particularly important if there is any chance that the full electrostatics of the emitter configuration is not properly understood.

## 7. Discussion

### *(a) Summary*

This paper has provided a more careful definition of the orthodox emission hypothesis, and has confirmed that—by using the concept of scaled barrier field—experimental tests for lack of field emission orthodoxy can be developed. After re-analyzing old-style CFE experiments, the limits involved in the range-ends test have been refined. Thus, after 90 years of practical CFE, we now have (for the first time) a simple, quantitative easy-to-apply test for determining when emission is not orthodox, and hence when FN-type equations certainly do not apply to a given emission situation as predictive/explanatory equations, but are serving just as empirical fitting equations.

The improved test has been applied to various types of field electron emitter, including many non-metal emitters and many LAFEs. Some exhibit apparently orthodox behaviour, others do not. Two-section "kinked" FN plots are characteristic of "saturation", which is usually due to series resistance in the current path . As might be expected, the high-$f$ section of such plots tested non-orthodox, in both cases tested. But in one case the low-$f$ section also tested non-orthodox. There are also single-section FN plots that tested non-orthodox.

Not surprisingly, field-dependent geometry and saturation effects appear as likely causes of non-orthodoxy. The finding that most carbon emitters tested, including some CNT-based emitters, have apparently orthodox behaviour was slightly unexpected. This perhaps suggests that when a CNT-based emission situation tests non-orthodox, this is most likely to be due to series resistance in the substrate or in the substrate-to-CNT contact.

*(b) General implications and future theoretical development*

Non-orthodoxy usually implies that a spurious FEF-value will be generated if orthodox (or simpler) data-analysis formulae are used. Hence, it is arguable that, as a routine part of data analysis, all new forms of field emitter should have their current-voltage characteristics tested for lack of orthodoxy, and that the results of the test should be included in related publications. A copy of the spreadsheet provided as part of electronic supplementary material could be used to do this.

There is also a case for applying the test to representative sets of past results, for the different LAFE types developed, to see if any systematic features of behaviour can be detected. One would certainly expect that the physical structure of an emitter, and the resistivities of the materials involved, would affect the extent to which the related CFE system is non-orthodox.

Lack of orthodoxy need not imply poor emitter quality—for example, in some applications the "ballasting" provided by series resistance is a practical advantage. What non-orthodoxy implies is that orthodox (and simpler) data-analysis methods are not applicable, and may generate spurious values for characterization parameters.

Clearly, there is a need for more sophisticated data-analysis theory. Although useful progress has been made in modelling series-resistance effects (e.g., Cha et al. 2006, Chen et al. 2010), this is only one of many potential causes of non-orthodoxy (as shown by the list of excluded complications in §1(c)). The way in which orthodox data-analysis theory needs to be generalized will depend, in part, on the particular cause of non-orthodoxy. The overall task seems likely to be highly complicated, will almost certainly need to proceed in stages, and will be addressed elsewhere in due course.

In conclusion, it is believed that the test developed here will prove an useful tool in developing field emission technology. For experimentalists it should help in assessing the validity of their extracted parameters; for technologists, it may assist with the choice of emitter material; for theoreticians it should help identify material situations that need detailed attention. At present, the test is a relatively coarse tool, aimed primarily at identifying the existence of problems. Hopefully, as results build up from its use, it may be possible to refine the test to become more informative.

More generally, it is hoped that the development of this test may help start a process of making the link between theory and experiment in field electron emission science stronger and more quantitative.

I thank the University of Surrey for provision of facilities.

**References**

Abbott, F. R. & Henderson, J. E. 1939 The range and validity of the field current equation. *Phys. Rev.* **56**, 113-118.

Barbour J. P., Dolan, W. W., Trolan, J. K., Martin, E. E. & Dyke W. P. 1953 Space-charge effects in field emission. *Phys. Rev.* **92**, 45-51.

Cha, S. I., Kim, K. T., Arshad, S. N., Mo, C. B., Lede, K. H. & Hong, S. H. 2006 Field emission behaviour of a carbon-nanotube-implanted Co nanocomposite fabricated from pearl-necklace-structured carbon nanotube/Co powders. *Adv. Mater.* **18**, 553-558.

Chen, L. F., Ji, Z. G., Mi, Y. H., Ni, H.L. & Zhao, H. F. 2010 Nonlinear characteristics of the Fowler-Nordheim plots of carbon nanotube field emission. *Phys. Scr*. **82**, 035602. (doi:10.1088/0031-8949/82/03/035602)

Deane, J. H. B. & Forbes, R. G. 2008 The formal derivation of an exact series expansion for the principal Schottky-Nordheim barrier function $v$, using the Gauss hypergeometric equation. *J. Phys. A: Math. Theor.* 41, 395301. (doi: 10.1088/1751-8113/41/39/395301)

Dyke, W. P & Dolan, W. W. 1956 Field emission *Adv. Electr. Electron Phys.* **8**, 89-185.

Dyke, W. P. & Trolan, J. K. Field emission: large current densities, space charge and the vacuum arc. *Phys. Rev.* **89**, 799-808.

Eyring, C. F., Mackeown, S. S. & Millikan R. A. 1928 Field currents from points. *Phys. Rev.* **31**, 900-909.

Forbes, R. G. 2008a Description of field emission current/voltage characteristics in terms of scaled barrier field values (*f*-values). *J. Vac. Sci. Technol. B* **26**, 209-213. (doi: 10.1116/1.2834563)

Forbes, R. G. 2008b Physics of generalized Fowler-Nordheim-type equations. *J. Vac. Sci. Technol. B* **26**, 788-793. (doi: 10.1116/1.2827505)

Forbes, R. G. 2008c On the need for a tunneling prefactor in Fowler-Nordheim tunnelling theory. *J. Appl. Phys*. **103***,* 114911 (doi: 10.1063/1.2937077)

Forbes, R. G. 2009 Use of Millikan-Lauritsen plots, rather than Fowler-Nordheim plots, to analyze field emission current-voltage data. *J. Appl. Phys.* **105***,* 114313 (doi: 10.1063/1.3140602)

Forbes, R. G. 2012a Extraction of emission parameters for large area field emitters using a technically complete Fowler-Nordheim-type equation. *Nanotechnology* **23**, 095706. (doi: 10.1088/0957-4484/23/9/095706)

Forbes, R. G. 2012b Comment on 'Metallic nanowire-graphene hybrid nanostructures for highly flexible field emission devices' *Nanotechnology* **23,** 28801 (doi: 10.1088/0957-4484/23/28/288001)

Forbes, R. G. & Deane J. H. B. 2007 Re-formulation of the standard theory of Fowler-Nordheim tunnelling and cold field electron emission. *Proc. R. Soc. Lond, A* **463**, 2907-2927. (doi: 10.1098/rspa.2007.0030)

Forbes, R. G. & Deane J. H. B. 2010 Comparison of approximations for the principal Schottky-Nordheim barrier function $v(f)$, and comments on Fowler-Nordheim plots *J. Vac. Sci. Technol. B* **28**, C2A33. (doi: 10.1116/1.3363858)

Forbes, R. G., Fischer, A. & Mousa, M. S. 2012 Use of a new type of intercept correction factor to improve Fowler-Nordheim plot analysis. Submitted for publication; arXiv:1208.3820.

Fowler, R. H. & Nordheim L. 1928 Electron emission in intense electric fields. *Proc. R. Soc. Lond. A* **119**, 173-181. (doi: 10.1098/rspa.1928.0091)

Good, R. H., jr, & Müller, E. W. 1956 Field emission, in: S. Flugge (ed.) *Encyclopedia of Physics, Vol. 21, Electron-emission gas discharges* (Springer-Verlag, Berlin), pp. 176-231.

Gomer, R. 1961 *Field emission and field ionization* (Harvard University Press, Cambridge, Mass).

Holbrow, C. H. 2003 In appreciation. Charles C. Lauritsen: A reasonable man in an unreasonable world. *Phys. perspect.* **5**, 419-472. See pp 426-428.

Landau, L. D. & Lifschitz, E. M. 1956 *Quantum Mechanics*. Oxford, UK: Pergamon. See equation (50.12).

Lauritsen, C. C. 1929 *Electron Emission from Metals in Intense Electric Fields.* PhD thesis, California Institute of Technology, 1929.

Lee, J., Lee, H., Heo, K., Lee, B. Y. & Hong, S. 2012 Reply to Comment on 'Metallic nanowire–graphene hybrid nanostructures for highly flexible field emission devices'. *Nanotechnology* **23**, 288002. (doi:10.1088/0957-4484/23/28/288002)

Li, J. L., Chen, M. M., Tian, S. B., Jin, A., Xia, X. X. & Gu C. Z. 2011 Single-crystal SnO2 nanoshuttles: shape-controlled synthesis, perfect flexibility and high-performance field emission *Nanotechnology* **22**, 505601. (doi:10.1088/0957-4484/22/50/505601)

Lilienfeld, J. E. 1922 The auto-electronic discharge and its application to the construction of a new form of X-ray tube. *Amer. J. Roentgenology* **9**, 172-179.

Lin, J., Huang, Y., Bando, Y., Tang, C.C., Li, C. & Golberg, D. 2010 Synthesis of $In_2O_3$ nanowire-decorated $Ga_2O_3$ nanobelt heterostructures and their electrical and field-emission properties. *ACS Nano* **4**, 2452-2458. (doi:10.1021/nn100254f)

Miller, H.C. 1967 Change in field intesification factor *β* of an electrode projection (whisker) at short gap lengths, *Phys. Rev.* **38**, 4501-4504.

Millikan, R. A. & Lauritsen, C. C. 1928 Relation of field-currents to thermionic-currents. *Proc. Natl. Acad. Sci. U.S.A.* **14**, 45-49.

Nordheim, L. W. 1928 The effect of the image force on the emission and reflexion of electrons by metals. *Proc. R. Soc. Lond. A* **121**, 626-639.

Qin, X.-Z., Wang, W.-L., Xu, N.-S., Li, Z.-B. & Forbes, R.G. 2010 Analytical treatment of cold field electron emission from a nanowall emitter, including quantum confinement effects. *Proc. R. Soc. Lond. A* **467**, 1029-1051. (doi: 10.1098/rspa.2010.0460)

Rother, F, 1926 Über den Austritt von Elektronen aus kalten Metallen. *Ann. Phys.* **20**, 317-372.

Saito, Y. (ed.) 2010 *Carbon nanotube and related field emitters* (Wiley-VCH, Weinheim).

Schottky, W. 1914 Über den Einfluss von Strukturwirkungen, besonders der Thomsonschen Bildkraft, auf die Elektronenemission der Metalle. *Physik. Zeitschr.* **15**, 872-878.

Spindt, C. A., Brodie, I., Humphrey L. & Westerberg, E. R. 1976 Physical properties of thin-film field emission cathodes. *J. Appl. Phys.* **47**, 5248-5263.

Stern, T. E., Gossling, B. S. & Fowler, R. H. 1929 Further studies in the emission of electrons from cold metals. *Proc. R. Soc. Lond. A* **124**, 699-723.

Xu, N.S. & Huq, E. 2005 Novel cold cathode materials. *Mat. Sci. Eng. Repts.* **R48**, 47-189.

Zhai, T. Y., Fang, X. S., Bando, Y., Dierre, B., Liu, B. D., Zeng, H. B., Xu, X. J., Huang, Y., Yuan, X. L., Sekiguchi, T. & Golberg, D. 2009 Characterisation, cathodiluminescence, and field emission properties of morphology-tunable CdS micro/nanostructures. *Adv. Funct. Mater.* **19**, 2423-2430.

Zhai, T. Y., Liang, L., Ma., Y., Liao, M. Y., Wang, X., Fang, X. S., Yao, J. N., Bando. Y. & Golberg, D. 2011 One-dimensional inorganic nanostructures: synthesis, field-emission and photodetection. *Chem. Soc. Rev.* **40**, 2986-3004.

Zhu, W. (ed.) 2001 *Vacuum Microelectronics* (Wiley, New York).

Žumer, M., Zajec, B., Rozman, R. & Nemanič V. 2012 Breakdown voltage reliability improvement in gas-discharge tube surge protectors employing graphite field emitters. *J. Appl. Phys.* **111**, 083301. (doi:10.1063/1.4704699)

**Table 1**

Table 1. Some universal constants used in field emission. Values are given in the eV-based units often used, in order to simplify calculations when work functions are given in eV and fields in V/nm, and current densities are needed in A/m$^2$.

| Name | Symbol | Expression | Numerical value | Units |
|---|---|---|---|---|
| First Fowler-Nordheim constant | $a$ | $e^3/8\pi h_P$ | 1.541 434 | µA eV V$^{-2}$ |
| Second Fowler-Nordheim constant | $b$ | $(4/3)(2m_e)^{1/2}/e\hbar$ | 6.830 890 | eV$^{-3/2}$ (V/nm) |
| Schottky constant | $c$ | $(e^3/4\pi\varepsilon_0)^{1/2}$ | 1.199 985 | eV (V/nm)$^{-1/2}$ |
| - | - | $c^{-2}$ | 0.694 4616 | eV$^{-2}$ (V/nm) |
| - | - | $bc^2$ | 9.836 238 | eV$^{1/2}$ |
| - | - | $ac^{-4}$ | 7.433 980×10$^{11}$ | A m$^{-2}$ eV$^{-3}$ |

**Table 2**

Table 2. (a) Values of the parameters $F_R$, $\eta$ and $\theta$ for a SN barrier of zero-field height $\phi$. (b) The $f$-values that set the "apparently reasonable" range $f_{\text{low.}} \leq f^{\text{extr}} \leq f_{\text{up.}}$, and the "clearly unreasonable" ranges $f^{\text{extr}} < f_{\text{lb}}$ and $f^{\text{extr}} > f_{\text{ub}}$.

| $\phi$ (eV) | $F_R$ (V/nm) | $\eta$ | $\theta$ (A/m$^2$) | $f_{\text{lb}}$ | $f_{\text{low.}}$ | $f_{\text{up.}}$ | $f_{\text{ub}}$ |
|---|---|---|---|---|---|---|---|
| 5.50 | 21.01 | 4.1942 | 1.24×10$^{14}$ | 0.09 | 0.14 | 0.41 | 0.69 |
| 5.00 | 17.36 | 4.3989 | 9.29×10$^{13}$ | 0.095 | 0.14 | 0.43 | 0.71 |
| 4.50 | 14.06 | 4.6368 | 6.77×10$^{13}$ | 0.10 | 0.15 | 0.45 | 0.75 |
| 4.00 | 11.11 | 4.9181 | 4.76×10$^{13}$ | 0.105 | 0.16 | 0.48 | 0.79 |
| 3.50 | 8.51 | 5.2577 | 3.19×10$^{13}$ | 0.11 | 0.17 | 0.51 | 0.85 |
| 3.00 | 6.25 | 5.6790 | 2.01×10$^{13}$ | 0.12 | 0.18 | 0.54 | 0.91 |
| 2.50 | 4.34 | 6.2210 | 1.16×10$^{13}$ | 0.13 | 0.20 | 0.59 | 0.98 |

**Table 3**

Table 3. Results of applying the improved range-ends test to selected old-style emission situations. With the exception of data set 10, which used steel, all emitters are tungsten. For the extracted *f*-values, the subscript meanings are: "low." & "up." = lower & upper ends of range used for steady emission; "nlin" = value at which non-linearity is observed; "pulslim" = highest value used in pulsed emission experiment. Entry 1 is the data set originally used (in Forbes 2012a,b) to establish range limits. Asterisks mark *f*-values that lie outside the revised "apparently physically reasonable" test limits $0.15 < f < 0.45$, but not outside the "clearly unreasonable" limits $f < 0.10$ and $f > 0.75$.

| No. | Year | Experimenters | $f_{\text{low.}}^{\text{extr}}$ | $f_{\text{up.}}^{\text{extr}}$ | $f_{\text{nlin.}}^{\text{extr}}$ | $f_{\text{pulslim.}}^{\text{extr}}$ |
|---|---|---|---|---|---|---|
| 1 | 1953 | Dyke & Trolan, X89 | 0.20 | 0.34 | 0.48 | 0.59 |
| 2 | 1972 | Young et al. | 0.19 | 0.28 | | |
| 3 | 1965 | Van Oostrom | 0.16 | 0.30 | | |
| 4 | 1956 | Barbour et al. | 0.19 | 0.24 | 0.58 | 0.58 |
| 5 | 1955 | Müller | 0.18 | 0.27 | | |
| 6 | 1953 | Dyke & Trolan, X81 | 0.21 | 0.35 | 0.44 | 0.69 |
| 7 | 1940 | Haefer | 0.39 | *0.59 | | |
| 8 | 1939 | Abbott & Henderson | *0.11 | 0.29 | | |
| 9 | 1936 | Müller | 0.25 | 0.37 | | |
| 10 | 1928 | Eyring et al. | 0.18 | 0.22 | | |
| 11 | 1928 | de Bruyne | 0.38 | *0.50 | | |
| 12 | 1928 | Millikan & Lauritsen | 0.22 | *0.46 | | |
| 13 | 1926 | Gossling & co-workers | 0.20 | 0.32 | | |
| 14 | 1926 | Millikan & Eyring | 0.18 | *0.52 | | |

**Table 4**

Table 4. Results of applying the improved range-ends test to various published Fowler-Nordheim plots, as described in the electronic supplementary material. A single asterisk indicates that the extracted *f*-value is outside the "apparently reasonable" range, a double asterisk that the value is "clearly unreasonable".

| No. | Material | $\gamma^{*1}$ | $\phi$ (eV) | $f_{\text{lower}}^{\text{extr}}$ | $f_{\text{upper}}^{\text{extr}}$ |
|---|---|---|---|---|---|
| 15 | Mo emitting cones, in Spindt array | | 4.60 | 0.36 | *0.73 |
| 16 | Au posts on flexible graphene film | | 5.10 | **2.81 | **5.21 |
| 17 | Flexible SnO$_2$ nanoshuttle, low-*f* section | 13000 | 4.70 | **5.60 | **33.2 |
| 18 | Flexible SnO$_2$ nanoshuttle, high-*f* section | | 4.70 | **0.76 | **1.03 |
| 19 | Amorphous carbon film | | 5.10 | 0.21 | *0.45 |
| 20 | Vertical carbon nanosheet | | 5.10 | 0.22 | 0.41 |
| 21 | Edges of flat graphite platelet | | 5.10 | 0.17 | 0.22 |
| 22 | CNT on tungsten needle | | 5.10 | 0.32 | 0.42 |
| 23 | Single clean SWNT | | 5.10 | *0.11 | 0.18 |
| 24 | CNT mat on silicon, low-*f* section | | 5.10 | 0.14 | 0.18 |
| 25 | CNT mat on silicon, high-*f* section | | 5.10 | **1.23 | **1.78 |
| 26 | CNT-in-polymner composite, 20% loading | | 5.00 | 0.15 | 0.23 |
| 27 | CNT-in-metallic-glass composite | | 4.60 | *0.54 | **0.87 |
| 28 | Graphene edge on ZnO nanowire tips | 10179 | 4.60 | **0.84 | **2.16 |
| 29 | ZnO nanorod array | 5750 | 4.45 | **0.87 | **1.66 |
| 30 | In$_2$O$_3$-decorated Ga$_2$O$_3$ nanobelt | 4002 | 4.80 | 0.20 | 0.38 |
| 31 | CdS nanotip array | 4819 | 4.20 | *0.68 | **3.98 |

# Development of a simple quantitative test for lack of field emission orthodoxy

## Richard G. Forbes

[Permanent e-mail alias: r.forbes@trinity.cantab.net]

## Electronic Supplementary Material (1)

### Origins of current-voltage data

This document provides further information about the origins of the current-voltage data used in preparing Tables 3 and 4 in the main paper, and notes how to convert plot slopes derived from older forms of data plot to the equivalent slope value for a Fowler-Nordheim plot made using natural logarithms. Details of the analysis of the Millikan-Lauritsen plots (Millikan & Lauritsen 1928, Forbes 2009)) and Fowler-Nordheim plots (Stern et al 1929) used in preparing the tables, and of the operation of the spreadsheet used, are given in the separate EXCEL-spreadsheet document OrthTest3.xls.

A copy of this spreadsheet can be used for the analysis of new FN and ML plots.

### Data plotting formats

In the mid-to-late 1920s, current-voltage ($i$-$V$) data relating to what was then called auto-electronic emission were presented either as a direct plot of $i$ against $V$, or—following the work of Schottky (1923)—in the form of a *Schottky plot*, i.e., a plot of $i$ against $V^{1/2}$. [Or in some equivalent form using a current density $J$ and/or a presumed local field $F$ proportional to $V$.]

It was shown in early 1928, by Lauritsen (Millikan & Lauritsen 1928), Oppenheimer (1928a,b) and Fowler & Nordheim (1928), that there were good experimental and theoretical reasons for making plots against $1/V$ (or, equivalently, $1/F$). Following this discovery, Millikan and his co-workers published re-presentations of much existing data, in the form of what I call "Millikan-Lauritsen (ML) plots", i.e., plots of $\ln\{i\}$ or $\log_{10}\{i\}$ against $1/V$ [or some equivalent form involving $J$ and/or $F$].

In the cases of early data, the plots analyzed in the present work are the re-presentations by Millikan and co-workers. It is indicated in Table 5 where this has been done.

The plotting method now called a Fowler-Nordheim (FN) plot (i.e., a plots of $\ln\{i/V^2\}$ or $\log_{10}\{i/V^2\}$ against $1/V$, or an equivalent form) was introduced by Stern et al. in 1929. The term "field electron emission" came into use as a replacement for "auto-electronic emission" in the 1930s, particularly in the German literature (e.g., Müller 1936, Haefer, 1940). The present author now prefers the term "cold field electron emission (CFE)".

Most experimental CFE data plots made in the first part of the 20th century used common logarithms rather than natural logarithms. For both ML and FN plots, the relationship between the slope $S^{(e)}$ of a plot made using natural logarithms and the slope $S^{(10)}$ of a plot made using common logarithms is:

$$S^{(e)} = \ln(10) \cdot S^{(10)} \approx 2.303 \cdot S^{(10)}. \tag{A1}$$

In the spreadsheet, the logathithm type ("e" or "10") is entered, and the spreadsheet makes this correction automatically.

As indicated in Table 5, many experimental CFE data plots made in the first part of the 20th century were ML plots rather than FN plots. A given set of experimental $i$-$V$ data can be plotted in either way.

For a given set of data, an approximate relationship between the magnitudes of the slope $S_{FN}$ (or $S_2$) of a FN plot of this data and the slope $S_{ML}$ (or $S_0$) of a ML plot of the data is (Forbes 2009):

$$|S_{FN}| = |S_{ML}| - 2 \cdot (1/X)_{mid}, \qquad (A2)$$

where $(1/X)_{mid}$ is the mid-range value of the range of values of $(1/X)$ used in the plotted data, where $X$ is the independent variable.

In the spreadsheet, the plot type ("FN") or ("ML") is entered, and the spreadsheet makes this correction automatically.

**Classical cold field electron emission data**

The origins of the classical CFE data used to prepare Table 3 in the main paper are shown in Table 5. Sets 12 and 14 refer to cylindrical wire emitters; the remaining sets to point emitters. Set 10 refers to a steel emitter; the remaining sets to tungsten emitters. In all cases, the work-function is taken as 4.50 eV.

Table 5. Origins of classical CFE data used to prepare Table 3.

| No. | Plot type | Year | Experimenters | Plot origin | Where plot used was published, if different |
|-----|-----------|------|---------------|-------------|---------------------------------------------|
| 1   | ML | 1953 | Dyke & Trolan, X89 | Fig.2 | |
| 2   | FN | 1972 | Young et al. | Fig.18 | |
| 3   | FN | 1965 | Van Oostrom | Fig.1 | |
| 4   | ML | 1956 | Barbour et al. | Fig.3 | |
| 5   | FN | 1955 | Müller | Fig.3 | |
| 6   | ML | 1953 | Dyke & Trolan, X81 | Fig.4 | |
| 7   | ML | 1940 | Haefer | Fig.4, "300 K" | |
| 8   | ML | 1939 | Abbott & Henderson | Fig.6 | |
| 9   | ML | 1937 | Müller | Fig.7, #13 | |
| 10  | ML | 1928 | Eyring et al. | Fig.4, #1 | |
| 11  | ML | 1928 | de Bruyne | Fig.6, "273°" | Millikan & Lauritsen 1929, Fig.2 |
| 12  | ML | 1928 | Millikan & Lauritsen | Fig.1 | |
| 13  | ML | 1926 | Gossling & co-workers | Fig.7, #6 | Eyring et al. 1929, Fig.5, #2 |
| 14  | ML | 1926 (?) | Millikan & Eyring (?) | unpublished | Millikan & Lauritsen 1928, Fig.2 |

# Modern field electron emission data

The origins of the modern data used to prepare Table 4 in the main paper are shown in Table 6. For the assumed work-function $\phi$, the standard literature value has been used, unless the original authors suggest a different value.

Table 6. Origins of modern data used to prepare Table 4.

| No. | Experimental article & page on which plot appears | Figure & plot references | Material | $\phi$ (eV) |
|---|---|---|---|---|
| 15 | Spindt et al. 1976, p5252 | Fig.8, 17-1-3G | Mo emitting cones, on silicon | 4.60 |
| 16 | Arif et al. 2011, p5 | Fig.3, 73 nm | Au posts on flexible graphene film | 5.10 |
| 17 | Li et al. 2011, p6 | Fig.7, nanoshuttle | Flexible $SnO_2$ nanoshuttle, low-$f$ section | 4.70 |
| 18 | Li et al. 2011, p6 | Fig.7, nanoshuttle | Flexible $SnO_2$ nanoshuttle, high-$f$ section | 4.70 |
| 19 | Tallin et al. 2001, p334 | Fig.4, (b) | Amorphous carbon film | 5.10 |
| 20 | Wang et al. 2006, p2 | Fig.2 | Vertical carbon nanosheet | 5.10 |
| 21 | Zumer et al. 2011, p5 | Fig.8, sample C | Edges of flat graphite platelet | 5.10 |
| 22 | de Jonge 2010, p71 | Fig.6.2(b) | CNT on tungsten needle | 5.10 |
| 23 | Dean 2010, p125 | Fig.10.2 | Single clean SWNT | 5.10 |
| 24 | Zhu et al. 2001, p269 | Fig.6.13, 48μm | CNT mat on silicon, low-$f$ section | 5.10 |
| 25 | Zhu et al. 2001, p269 | Fig.6.13, 48μm, | CNT mat on silicon, high-$f$ section | 5.10 |
| 26 | Connelly et al. 2009, p828 | Fig.5, 15.6% | CNT-in-polymer composite, 15.6% loading | 5.00 |
| 27 | Hojati-Talemi et al., 2011, p3 | Fig.2, 10% | CNT-in-metallic-glass composite | 4.60 |
| 28 | Maiti et al. 2011, p6 | Fig.6, G30 | Graphene edge on ZnO nanowire tips | 4.60 |
| 29 | Zhai et al. 2010, p2989 | Fig.4(d) | ZnO nanorod array | 4.45 |
| 30 | Lin et al. 2011, p2456 | Fig.6, 740μm | $InO_3$-decorated $Ga_2O_3$ nanobelt | 4.80 |
| 31 | Zhai et al. 2009, p2427 | Fig8(b) | CdS nanotips | 4.20 |

# References

Abbott, F. R. & Henderson, J. E. 1939 The range and validity of the field current equation. *Phys. Rev.* **56**, 113-118.

Arif, M., Heo, K., Lee, B. Y., Lee. J., Seo, D. H., Seo, S., Jian, J. and Hong, S. 2011 Metallic nanowire-graphene hybrid nanostructures for highly flexible field emission devices. *Nanotechnology* **22,** 355709 (doi:10.1088/0957-4484/22/35/355709)

Barbour J. P., Dolan, W. W., Trolan, J. K., Martin, E. E. & Dyke W. P. 1953 Space-charge effects in field emission. *Phys. Rev.* **92**, 45-51.

Connolly, T., Smith, R. C., Hernandez, Y., Gun'ko, Y., Coleman, J. N. & Carey, J. D. 2009 Carbon-nanotube–polymer nanocomposites for field-emission cathodes. *Small* **5**, 826–831. (doi: 10.1002/smll.200801094)

de Bruyne, N. A. 1928 The action of strong electric fields on the current from a thermionic cathode *Proc. R. Soc. Lond. A.* **52**, 423-437.

de Jonge, N. 2010 The optical performance of carbon nanotube field emitters. Chap. 6 in Saito 2010.

Dean, K. A. 2010 Field emission from single-wall nanotubes. Chap. 10 in Saito 2010.

Dyke, W. P. & Trolan, J. K. Field emission: large current densities, space charge and the vacuum arc. *Phys. Rev.* **89**, 799-808.

Eyring, C. F., Mackeown, S. S. & Millikan R. A. 1928 Field currents from points. *Phys. Rev.* **31**, 900-909.

Forbes, R. G. 2009 Use of Millikan-Lauritsen plots, rather than Fowler-Nordheim plots, to analyze field emission current-voltage data. *J. Appl. Phys.* **105**, 114313 (doi: 10.1063/1.3140602)

Gossling, B. S. and the research staff of the General Electric Co. Ltd. 1926 *Phil Mag.* **7**, 609-635.

Haefer, R. 1940 Experimentelle Untersuchungen zur Prüfung der wellenmechanischen Theorie der Feldelektronenemission. *Z. Phys.* **116**, 604-623.

Hojati-Talemi, P., Gibson, M. A., East, D. & Simon, G. P. 2011 High performance bulk metallic glass/carbon nanotube composite cathodes for electron field emission. *Appl. Phys. Lett.* **99**, 194104. (doi:10.1063/1.3659898)

Li, J. L., Chen, M. M., Tian, S. B., Jin, A., Xia, X. X. & Gu, C. Z. 2011 Single-crystal $SnO_2$ nanoshuttles: shape-controlled synthesis, perfect flexibility and high-performance field emission *Nanotechnology* **22**, 505601. (doi:10.1088/0957-4484/22/50/505601)

Lin, J., Huang, Y., Bando, Y., Tang, C.C., Li, C. & Golberg, D. 2010 Synthesis of $In_2O_3$ nanowire-decorated $Ga_2O_3$ nanobelt heterostructures and their electrical and field-emission properties. *ACS Nano* **4**, 2452-2458. (doi:10.1021/nn100254f)

Maiti, U. N., Maiti, S., Majumder, T. P. & Chattopadhyay, K. K. 2011 Ultra-thin graphene edges at the nanowire tips: a cascade cold cathode with two-stage field amplification. *Nanotechnology* **22**, 505703. (doi:10.1088/0957-4484/22/50/505703)

Millikan, R. A. & Eyring, C. F 1926 Laws governing the pulling of electrons out of metals by intense electrical fields. *Phys. Rev.* **27**, 51-67.

Millikan, R. A. & Lauritsen, C. C. 1928 Relation of field-currents to thermionic-currents. *Proc. Natl. Acad. Sci. U.S.A.* **14**, 45-49.

Millikan, R. A. & Lauritsen, C. C. 1929 Dependence of electron emission from metals upon field strengths and temperatures. *Phys. Rev.* **33**, 598-604.

Müller, E. W. 1937 Die Abhängigkeit der Feldelektronenemission von der Austrittsarbeit. *Z. Phys.* **102**, 734-761.

Müller, E. W. 1955 Work function of tungsten single crystal planes measured by the field emission microscope. *J. Appl. Phys.* **26**, 732-737

Oppenheimer, J. R. 1928a Three notes on the quantum theory of aperiodic effects. *Phys. Rev.* **13**, 66-81.

Oppenheimer, J. R. 1928b On the quantum theory of the autoelectronic field currents. *Proc. Natl. Acad. Sci. U.S.A.* **14**, 363-365.

Schottky, W. 1923 Über kalte und warme Electronentladungen. *Z. Phys*. **14**, 63-106.

Spindt, C. A., Brodie, I., Humphrey L. & Westerberg, E. R. 1976 Physical properties of thin-film field emission cathodes. *J. Appl. Phys.* **47**, 5248-5263.

Stern, T. E., Gossling, B. S. & Fowler, R. H. 1929 Further studies in the emission of electrons from cold metals. *Proc. R. Soc. Lond. A* **124**, 699-723.

Tallin, A.A., Coll, B. F., Menu, E. P., Markham, J. & Jaskie J. E. 1998 Carbon cathode requirements and emission characterization for low-voltage field emission displays. In: Silva, S. R. P., Robertson, J. Milne, W. I. & Amaratunga G. A. J. (Eds) *Amorphous Carbon: State of the Art* (World Scientific, Singapore).

Van Oostrom, A. G. J. 1965 *Validity of the Fowler-Nordheim model for field electron emission*. PhD thesis, University of Amsterdam; published as *Philips Res. Repts, Supplements 1966*, No. 1.

Wang, S., Wang, J. J., Miraldo, P., Zhu, M. Y., Outlaw, R., Hou, K., Zhao, X., Holloway, B. C., Manos, D., Tyler, T., Shenderova, O., Ray, M., Dalton, J. & McGuire, G. 2006 High field emission reproducibility and stability of carbon nanosheets and nanosheet-based backgated triode emission devices. *Appl. Phys. Lett.* **89**, 183103. (doi: 10.1063/1.2372708)

Young, R., Ward, J. & Scire, F. 1972 The topografiner: an instrument for measuring surface microtopography. *Rev. Sci. Instr*. **43,** 999-1011.

Zhai, T. Y., Fang, X. S., Bando, Y., Dierre, B., Liu, B. D., Zeng, H. B., Xu, X. J., Huang, Y., Yuan, X. L., Sekiguchi, T. & Golberg, D. 2009 Characterisation, cathodiluminescence, and field emission properties of morphology-tunable CdS micro/nanostructures. *Adv. Funct. Mater.* **19**, 2423-2430. (doi:10.1002/adfm.200900295)

Zhai, T. Y., Liang, L., Ma., Y., Liao, M. Y., Wang, X., Fang, X. S., Yao, J. N., Bando. Y. & Golberg, D. 2011 One-dimensional inorganic nanostructures: synthesis, field-emission and photodetection. *Chem. Soc. Rev.* **40**, 2986-3004. (doi: 10.1002/adfm.200900295).

Zhu, W., Baumann, P. K. & Bower, C. A. 2001 Novel cold cathode materials. Chap. 6 in Zhu 2001.

Zhu, W. (ed.) 2001 *Vacuum Microelectronics* (Wiley, New York).

Žumer, M., Zajec, B., Rozman, R. & Nemanič V. 2012 Breakdown voltage reliability improvement in gas-discharge tube surge protectors employing graphite field emitters. *J. Appl. Phys.* **111**, 083301. (doi:10.1063/1.470469).

**Development of a simple quantitative test for lack of field emission orthodoxy**
Richard G. Forbes
[Permanent e-mail alias: r.forbes@trinity.cantab.net]

<u>Electronic Supplementary Material (2)</u>

**SPREADSHEET FOR APPLYING ORTHODOXY TEST TO A FOWLER-NORDHEIM OR MILLIKAN-LAURITSEN PLOT**

This spreadsheet is protected, but no password has been used.
Colleagues are very welcome to make an unprotected copy of this spreadsheet in order to analyze results,
on the understanding that it is provided "as is", and that RGF takes no responsibility for its use by others.

**Constants**
b*c*c    9.836238
s(f_t)   0.950

| Information inputs | | | | | | | Data inputs | | | | | | Calculations | | | | | | | | | | Outputs | | | | | |
|---|---|---|---|---|---|---|---|---|---|---|---|---|---|---|---|---|---|---|---|---|---|---|---|---|---|---|---|---|
| No. | | phi | eta | FN? | ln? | | (1/X) lp | (1/X) nl | (1/X) ls | L ls | (1/X) hs | L hs | (−S)fit plot | (−S)fit Np | (1/X) mid | Corr to −S | (−S) corrected | Extrac const | 1/f lp | 1/f nl | 1/f ls | 1/f hs | f ls | f hs | f nl | f hp | No. | Notes |
| *Table 3* | | | | | | | | | | | | | | | | | | | | | | | | | | | | | |
| 1 | Dyke & Trolan, X89 | 4.50 | 4.637 | ML | 2 | e | 1.000 | 1.42 | 1.73 | 2.459 | | 4.157 | | 5.86 | 5.86 | 3.308 | −0.60 | 5.26 | 1.193 | 1.7 | 2.1 | 2.9 | 5.0 | 0.20 | 0.34 | 0.48 | 0.59 | 1 | a |
| 2 | Young et al. | 4.50 | 4.637 | FN | 0 | 10 | 2.303 | | | 2.500 | 4.78 | 3.000 | 1.27 | 2.70 | 6.22 | 3.150 | 0.00 | 6.22 | 1.411 | | | 3.5 | 5.4 | 0.19 | 0.28 | | | 2 | |
| 3 | Van Oostrom | 4.50 | 4.637 | FN | 0 | 10 | 2.303 | | | 4.000 | | 7.500 | | 1.59 | 3.65 | 5.750 | 0.00 | 3.65 | 0.829 | | | 3.3 | 6.2 | 0.16 | 0.30 | | | 3 | b |
| 4 | Barbour et al. | 4.50 | 4.637 | ML | 2 | e | 1.000 | − | 0.50 | 1.200 | −14.00 | 1.500 | −19.00 | 16.67 | 16.67 | 1.350 | −1.48 | 15.19 | 3.447 | − | 1.7 | 4.1 | 5.2 | 0.19 | 0.24 | 0.58 | − | 4 | |
| 5 | Müller, 1955 | 4.50 | 4.637 | FN | 0 | 10 | 2.303 | | | 2.600 | −11.60 | 3.980 | −15.30 | 2.68 | 6.17 | 3.290 | 0.00 | 6.17 | 1.401 | | | 3.6 | 5.6 | 0.18 | 0.27 | | | 5 | |
| 6 | Dyke & Trolan, X81 | 4.50 | 4.637 | ML | 2 | e | 1.000 | 1.43 | 2.25 | 2.800 | −9.00 | 4.750 | −18.80 | 5.03 | 5.03 | 3.775 | −0.53 | 4.50 | 1.021 | 1.5 | 2.3 | 2.9 | 4.8 | 0.21 | 0.35 | 0.44 | 0.69 | 6 | |
| 7 | Haefer | 4.50 | 4.637 | ML | 2 | 10 | 2.303 | | | 6.100 | −5.35 | 9.300 | −7.400 | 0.64 | 1.48 | 7.700 | −0.26 | 1.22 | 0.276 | | | 1.7 | 2.6 | 0.39 | 0.59 | | | 7 | |
| 8 | Abbott & Henderson | 4.50 | 4.637 | ML | 2 | 10 | 2.303 | | | 3.900 | −5.00 | 10.000 | −16.10 | 1.82 | 4.19 | 6.950 | −0.29 | 3.90 | 0.886 | | | 3.5 | 8.9 | 0.11 | 0.29 | (d) | | 8 | |
| 9 | Müller, 1937 | 4.50 | 4.637 | ML | 2 | 10 | 2.303 | | | 0.165 | −4.20 | 0.246 | −7.000 | 34.50 | 79.44 | 0.206 | 8.13 | 71.31 | 16.188 | | | 2.7 | 4.0 | 0.25 | 0.37 | | | 9 | c |
| 10 | Eyring et al. (steel) | 4.50 | 4.637 | ML | 2 | 10 | 2.303 | | | 5.250 | −8.75 | 6.375 | −10.75 | 1.78 | 4.09 | 5.813 | −0.34 | 3.75 | 0.851 | | | 4.5 | 5.4 | 0.18 | 0.22 | | | 10 | |
| 11 | de Bruyne | 4.50 | 4.637 | ML | 2 | 10 | 2.303 | | | 3.190 | 3.40 | 4.160 | 2.00 | 1.44 | 3.32 | 3.675 | −0.54 | 2.78 | 0.631 | | | 2.0 | 2.6 | 0.38 | 0.50 | | | 11 | |
| 12 | Lauritsen | 4.50 | 4.637 | ML | 2 | 10 | 2.303 | | | 1.220 | −4.00 | 2.500 | −9.00 | 3.91 | 8.99 | 1.860 | −1.08 | 7.92 | 1.798 | | | 2.2 | 4.5 | 0.22 | 0.46 | | | 12 | |
| 13 | Gossling & co-workers | 4.50 | 4.637 | ML | 2 | 10 | 2.303 | | | 0.525 | −3.40 | 0.845 | −8.00 | 12.50 | 28.78 | 0.685 | −2.92 | 25.86 | 5.871 | | | 3.1 | 5.0 | 0.20 | 0.32 | | | 13 | |
| 14 | Millikan lab | 4.50 | 4.637 | ML | 2 | 10 | 2.303 | | | 2.000 | −3.40 | 5.700 | −11.00 | 2.05 | 4.73 | 3.850 | −0.52 | 4.21 | 0.956 | | | 1.9 | 5.4 | 0.18 | 0.52 | | | 14 | |
| *Table 4* | | | | | | | | | | | | | | | | | | | | | | | | | | | | | |
| 15 | Mo emitting cones (Spindt array) | 4.60 | 4.586 | FN | 0 | 10 | 2.303 | | | 5.200 | −7.59 | 10.600 | −10.28 | 0.50 | 1.15 | 7.900 | 0.00 | 1.15 | 0.263 | | | 1.4 | 2.8 | 0.36 | 0.73 | | | 15 | |
| 16 | Au posts on flexible graphene film | 5.10 | 4.356 | FN | 0 | e | 1.000 | | | 0.332 | | 0.624 | | | 2.36 | 0.478 | 0.00 | 2.36 | 0.570 | | | 0.2 | 0.4 | 2.81 | 5.29 | (e) | | 16 | b |
| 17 | Flexible SnO2 nanoshuttles, low f | 4.70 | 4.537 | FN | 0 | e | 1.000 | | | 0.337 | −3.19 | 2.000 | −3.83 | 0.38 | 0.38 | 1.169 | 0.00 | 0.38 | 0.089 | | | 0.0 | 0.2 | 5.60 | 33.23 | | | 17 | |
| 18 | Flexible SnO2 nanoshuttles, high f | 4.70 | 4.537 | FN | 0 | e | 1.000 | | | 0.244 | −1.71 | 0.330 | −3.19 | 17.21 | 17.21 | 0.287 | 0.00 | 17.21 | 3.993 | | | 1.0 | 1.3 | 0.76 | 1.03 | | | 18 | |
| 19 | Amorphous carbon film | 5.10 | 4.356 | FN | 0 | e | 1.000 | | | 0.100 | −12.50 | 0.213 | −22.60 | 92.00 | 92.00 | 0.156 | 0.00 | 92.00 | 22.234 | | | 2.2 | 4.7 | 0.21 | 0.45 | | | 19 | |
| 20 | Vertical carbon nanosheet | 5.10 | 4.356 | FN | 0 | e | 1.000 | | | 2.000 | −9.95 | 3.780 | −18.85 | 5.00 | 5.00 | 2.890 | 0.00 | 5.00 | 1.208 | | | 2.4 | 4.6 | 0.22 | 0.41 | | | 20 | |
| 21 | Flat graphite platelet edges | 5.10 | 4.356 | ML | 2 | 10 | 2.303 | | | 2.500 | −7.62 | 3.333 | −10.58 | 3.55 | 8.18 | 2.917 | −0.69 | 7.49 | 1.811 | | | 4.5 | 6.0 | 0.17 | 0.22 | | | 21 | |
| 22 | CNT on tungsten STFE | 5.10 | 4.356 | FN | 0 | e | 1.000 | | | 3.020 | | 4.000 | | | 3.244 | 3.510 | 0.00 | 3.24 | 0.784 | | | 2.3 | 3.1 | 0.32 | 0.42 | | | 22 | b |
| 23 | Single clean SWCNT | 5.10 | 4.356 | FN | 0 | 10 | 2.303 | | | 4.150 | −12.50 | 6.900 | −15.00 | 2.36 | 5.43 | 5.525 | 0.00 | 5.43 | 1.312 | | | 5.4 | 9.1 | 0.11 | 0.10 | | | 23 | |
| 24 | CNT mat on Si, low-f regime | 5.10 | 4.356 | FN | 0 | e | 1.000 | | | 4.270 | −24.82 | 5.440 | −31.18 | 5.44 | 5.44 | 4.855 | 0.00 | 5.44 | 1.314 | | | 5.6 | 7.1 | 0.14 | 0.18 | | | 24 | |
| 25 | CNT mat on Si, high-f regime | 5.10 | 4.356 | FN | 0 | e | 1.000 | | | 2.900 | −23.36 | 4.270 | −24.46 | 0.80 | 0.80 | 3.585 | 0.00 | 0.80 | 0.194 | | | 0.6 | 0.8 | 1.21 | 1.78 | | | 25 | |
| 26 | CNT-in-polymer composite | 5.00 | 4.399 | FN | 0 | e | 1.000 | | | 0.180 | −15.56 | 0.280 | −25.66 | 101.00 | 101.00 | 0.230 | 0.00 | 101.00 | 24.169 | | | 4.4 | 6.8 | 0.15 | 0.23 | | | 26 | |
| 27 | CNT-in-metallic-glass composite | 4.60 | 4.586 | FN | 0 | e | 1.000 | | | 0.502 | −8.93 | 0.808 | −12.00 | 10.03 | 10.03 | 0.655 | 0.00 | 10.03 | 2.303 | | | 1.2 | 1.9 | 0.54 | 0.87 | | | 27 | |
| 28 | Graphene edges on ZnO nanowire tip | 4.60 | 4.586 | FN | 0 | e | 1.000 | | | 0.258 | −2.93 | 0.663 | −6.10 | 7.83 | 7.83 | 0.461 | 0.00 | 7.83 | 1.797 | | | 0.5 | 1.2 | 0.84 | 2.16 | | | 28 | |
| 29 | ZnO nanrod array | 4.45 | 4.663 | FN | 0 | e | 1.000 | | | 0.200 | −2.80 | 0.384 | −5.25 | 13.32 | 13.32 | 0.292 | 0.00 | 13.32 | 3.006 | | | 0.6 | 1.2 | 0.87 | 1.66 | | | 29 | |
| 30 | InO3 nanowire decorated Ga2O3 nanobelt | 4.80 | 4.490 | FN | 0 | e | 1.000 | | | 0.667 | −3.67 | 1.253 | −13.55 | 16.86 | 16.86 | 0.960 | 0.00 | 16.86 | 3.953 | | | 2.6 | 5.0 | 0.20 | 0.38 | | | 30 | |
| 31 | CdS nanotips | 4.20 | 4.800 | FN | 0 | e | 1.000 | | | 0.090 | 0.60 | 0.530 | −5.00 | 12.73 | 12.73 | 0.310 | 0.00 | 12.73 | 2.791 | | | 0.3 | 1.5 | 0.68 | 3.98 | | | 31 | |
| A | B | C | D | E | F | G | H | I | J | K | L | M | N | O | P | Q | R | S | T | U | V | W | X | Y | Z | AA | AB | AC | AD |

**Operation of Spreadsheet, by columns**

- A    Sequence number.
- B    ENTER plot identification, if wished.
- C    ENTER assumed work-function value $\phi$.
- D    Parameter $\eta$ calculated from $\eta = b*c*c/\sqrt{\phi}$) [equation (10) in paper]
- E    ENTER "FN" if Fowler-Nordheim plot, ENTER "ML" if Millikan-Lauritsen plot
- F    Spreadsheet inserts "2" if ML entered in E, oherwise inserts "0"; these numbers are used in column S.
- G    ENTER "e" if vertical plot axis uses natural logarithms, ENTER "10" if vertical plot axis uses common logarithms.
- H    Spreadsheet inserts value of ln(10) if "10" entered in G, otherwise inserts 1; these numbers are used in column P.
- I    If relevant, ENTER (1/$X$) numerical value corresponding to highest pulsed value of independent variable (lowest pulsed value of 1/$X$) [see note (a) below].
- J    If relevant, ENTER (1/$X$) numerical value corresponding to value of independent variable at which non-linearity sets in [see note (a) below].
- K    ENTER (1/$X$) numerical value corresponding to left-hand (high-$f$) end of range of data points [see note (a) below].
- L    Enter logarithm value ($L$-value) corresponding to leftmost data-point, *as given by line fitted to data points*.
- M    ENTER (1/$X$) numerical value corresponding to right-hand (low-$f$) end of range of data points [see note (a) below].
- N    Enter logarithm value ($L$-value) corresponding to rightmost data-point, *as given by line fitted to data points*.
- O    Spreadsheet uses entries K to N to calculates slope $S^\wedge$fit of line fitted to data points. (ALTERNATIVELY, enter slope, taking care to get units correct.)
- P    If necessary, spreadsheet converts slope-value to eqivalent value for a plot using natural logarithms.
- Q    Spreadsheet calculates mid-range value (1/$X$_mid) of 1/$X$.
- R    For Millikan-Lauritsen plots, spreadsheet calculates correction to slope *magnitude* (to make result equivalent to FN plot) by applying formula: *correction* = – 2 / (1/$X$_mid).
- S    Column S applies correction if necessary and records effective value to be taken for magnitude of FN plot slope.
- T    Spreadsheet calculates extraction constant (for conversion from 1/$X$ to 1/$f$ for the plot in question) [see equation (27) in paper] using the formula: *const*= –$\eta$*s t/$S^\wedge$fit.
- U    Shows 1/$f$ value corresponding to entry in column I, if any.
- V    Shows 1/$f$ value corresponding to entry in column J, if any.
- W    Shows 1/$f$ value corresponding to entry in column K.
- X    Shows 1/$f$ value corresponding to entry in column M.
- Y    Shows $f$-value corresponding to low side of extracted range.
- Z    Shows $f$-value corresponding to low side of extracted range.
- AA    Shows $f$-value corresponding to onset of non-linearity (if entry present in column J).
- AB    Shows $f$-value corresponding to maximum pulsed field value (if entry present in column I).
- AC    Copy of sequence number.
- AD    Notes

**Notes**

(a) Powers of 10 on the horizontal axis can be neglected since these cancel out when evaluating $S/X$. But if a slope value is then entered directly, this must be done consistently.
(b) A slope value highlighted in green is a value given by original author; this value has been used in the calculations.
(c) Values of 1/$X$ have been calculated from $L$-values extracted from data plot, using Muller's tabulated information that: $L$ = 1.5 – 34.5*{1/$X$}, for his plot 13.
(d) Values highlighted in orange are in the "further investigation needed" range of extracted $f$-values.
(e) Values highlighted in red are in the "clearly unreasonable" range of extracted $f$-values.